\documentclass[twocolumn]{aastex631}
\usepackage{makecell}
\usepackage{tikz}
\usetikzlibrary{shapes.geometric, arrows}
\usepackage{physics}
\usepackage{amsmath}
\usepackage{nicefrac}
\usepackage{threeparttable}
\usepackage{subfigure}
\usepackage{float}
\usetikzlibrary{calc}

\shorttitle{StarShot}
\shortauthors{Tse et al.}

\newcommand{\A}[1]{\textcolor{blue}{#1}}

\graphicspath{{./}{figures/}}
\begin{document}
\title{Time-independent Simulations of Steady-State Accretion with Nuclear Burning}

\author[00000-0002-7604-1517]{Kaho Tse}
\affiliation{School of Physics and Astronomy, Monash University, VIC 3800, Australia}
\affiliation{OzGrav: The ARC Centre of Excellence for Gravitational Wave Discovery, Australia}
\affiliation{Joint Institute for Nuclear Astrophysics - Center for the Evolution of the Elements (JINA-CEE), Monash University, VIC 3800, Australia}

\author[0000-0002-3684-1325]{Alexander Heger}
\affiliation{School of Physics and Astronomy, Monash University, VIC 3800, Australia}
\affiliation{OzGrav: The ARC Centre of Excellence for Gravitational Wave Discovery, Australia}
\affiliation{Joint Institute for Nuclear Astrophysics - Center for the Evolution of the Elements (JINA-CEE), Monash University, VIC 3800, Australia}
\affiliation{ARC Centre of Excellence for Astrophysics in Three Dimensions (ASTRO-3D), Australia}

\author[0000-0002-8032-8174]{Ryosuke Hirai}
\affiliation{School of Physics and Astronomy, Monash University, VIC 3800, Australia}
\affiliation{OzGrav: The ARC Centre of Excellence for Gravitational Wave Discovery, Australia}
\affiliation{Joint Institute for Nuclear Astrophysics - Center for the Evolution of the Elements (JINA-CEE), Monash University, VIC 3800, Australia}

\author[0000-0002-6558-5121]{Duncan K. Galloway}
\affiliation{School of Physics and Astronomy, Monash University, VIC 3800, Australia}
\affiliation{OzGrav: The ARC Centre of Excellence for Gravitational Wave Discovery, Australia}
\affiliation{Joint Institute for Nuclear Astrophysics - Center for the Evolution of the Elements (JINA-CEE), Monash University, VIC 3800, Australia}
\affiliation{Institute for Globally Distributed Open Research and Education (IGDORE)}
\begin{abstract}
We construct a new formulation that allows efficient exploration of steady-state accretion processes onto compact objects.  Accretion onto compact objects is a common scenario in astronomy.  These systems serve as laboratories to probe the nuclear burning of the accreted matter.  Conventional stellar evolution codes have been developed to simulate in detail the nuclear reactions on the compact objects. In order to follow the case of steady burning, however, using these codes can be very expensive as they are designed to follow a time-dependent problem.  Here we introduce our new code \textsc{StarShot}, which resolves the structure of the compact objects for the case of stable thermonuclear burning, and is able to follow all nuclear species using an adaptive nuclear reaction network and adaptive zoning.  Compared to dynamical codes, the governing equations can  be reduced to time-independent forms under the assumption of steady-state accretion.  We show an application to accreting low mass X-ray binaries (LMXBs) with accretion onto a neutron-star as compact object.   The computational efficiency of \textsc{StarShot} allows us to explore the parameter space for stable burning regimes, and can be used to generate initial conditions for time-dependent evolution models.
\end{abstract}


\section{Introduction} \label{sec:intro}
In low-mass X-ray binaries (LMXBs),  compact objects accrete material from their companion stars, which have masses of $\lesssim 1~M_{\odot}$, due to Roche-Lobe overflow.  Type I (thermonuclear) X-ray bursts can occur when the accumulated fuel becomes hot and dense enough to encounter thermonuclear runaway.  They are the most frequent natural thermonuclear explosions in the Universe.  The first observations of these nuclear flashes were made by \citet{Grindlay1976} and \citet{Belian1976} in the 1970s.  Since then, the number of known burst sources has grown significantly.  Nowadays, more than seven thousand examples can be found in multi-instrument catalogues \citep[e.g.][]{minbarpaper}. They show a variety of properties in terms of duration, recurrence time, and energy released.

Episodes of stable nuclear burning, during which no bursts are observed, is likely to occur at high base heating, which is the amount of heating from the neutron star crust, as well as at high accretion rates \citep{keek2016}. It is because the temperature-dependence of nuclear reactions saturates at high temperature, and therefore energy generation is compensated by radiative and neutrino cooling \citep{Bildsten_1998}.  The thermal properties of neutron star envelopes are therefore favourable of quenching the bursts by stable burning.  The cooling phase of neutron stars post-outburst also provides opportunities to probe the thermal properties and compositions of the neutron star crust and ocean \citep[e.g.][]{brown2009}.  The large observational database implies that we do have lots of samples to compare to the developed theory.

Simulations of the thermonuclear burning on neutron star surface have also been developed over decades in an attempt to reproduce the behaviour of Type I X-ray bursts. Initially, \citet{paczynski1983} proposed the idea of modelling the bursts using a one-zone model, in which only a thin shell of fuel layer is considered for nuclear energy generation and cooling with radiative diffusion. The simulations have later been developed to be one-dimensional multi-zones, with larger nuclear networks with up to 1300 isotopes being adopted for nuclear energy generation \citep{woosley2004}. The simulation networks are continuously refined with more recent nuclear experimental data being updated. 

Time evolution codes for such simulations, for example, \textsc{Kepler} \citep{1978ApJ...225.1021W,2004ApJS..151...75W} and \textsc{MESA} \citep{mesa1,mesa2}, are able to reproduce different key features of bursts. For instance, the recurrence time of thermonuclear flashes \citep{heger2007b,Lampe_2016,Meisel_2018}, the detonation precursor of the bursts \citep{keek2011}, the quasi periodic oscillations in milli-hertz frequency domain (mHz QPOs) which are believed to be a special mode of nuclear burning \citep{heger_qpo,zamfir2014}, and the much longer timescale carbon flashes (known as `superbursts'; e.g.,  \citealt{keek2012}), which typically last for hours compared to Type I X-ray bursts in minutes.  The models are still able to achieve these burst properties even though the macroscopic effects, such as magnetic field and angular momentum transport, are not being considered in these simulations. 

Whereas dynamical codes may be ideal for studying the time-dependent unstable nuclear burning, due to their significant computational costs, they are not efficient to simulate steady-state burning.  In dynamical simulations for the binary system, the neutron star envelope is treated as a grid of Lagrangian zones in the radial direction.  To follow the accretion, for each time step, the structure of the entire grid has to be computed, including energy transport and nuclear reactions for all zones.  The evolution time step for the entire grid is restricted to follow the zone with the smallest time step.  In the case of the stationary structure of accreting neutron star envelop under stable burning, however, this coupling is unnecessary.

In light of this issue, we introduce our \textsc{StarShot} code for steady-state accretion models.  It makes use of the fact that the simulations of the entire grid can be decoupled for stationary solutions, which greatly reduces the computational time.  In Section~\ref{sec:style}, we show the details of solving the set of governing equations.  The algorithm of calculations for each generic time step is summarised in a flowchart in the same section, followed by the results in Section~\ref{sec:results}. In the final section, we go through the applications of the code and its possible extensions in future.

%They end up with costing unnecessarily computational time, which should actually be avoided when solving steady-state solutions.

%While both codes are able to provide modellings for the theoretical development of the neutron star structure, they inevitably require significant computational time for initial modelling setups, where in some cases are just the beginning of our scientific interests. The initial transient phase could have chains of bursts, which requires expensive calculations of materials and heat diffusion, as well as nuclear reactions.

\section{The \textsc{StarShot} code} \label{sec:style}
The basic differential equations that govern the accretion flow onto neutron stars include hydrodynamic equations, energy conservation and change of composition with time, expressed in Lagrangian coordinates as
\begin{align}
        \pdv{P}{m} &= -\frac{Gm}{4\pi r^4} - \frac{1}{4\pi r^2}\frac{\partial^2r}{\partial{}t^2}\;, \label{eq:momentum}\\
    \pdv{l}{m} &= \epsilon_n - \epsilon_v - \pdv{u}{t} - P\pdv{v}{t}\;, \\
    \pdv{X_j}{t} &= F(T, \rho, X_j)\;,
    \label{tradition_eqs}
\end{align}
where $P$ is the pressure; $m$ is the mass coordinate (mass interior to the current location); $G$ is the gravitational constant; $r$ is the radius; $l$ is the local luminosity; $\epsilon_n$ is the specific nuclear energy generation rate; $\epsilon_v$ is the specific neutrino loss rate; $u$ is the internal energy of gas; $v$ is the specific volume; $T$ is the temperature; $\rho$ is the density; and $X_j$ is the mass fraction of species $j$.  The function $F$ denotes the rate of change of the nuclear species due to nuclear reactions as a function of the current composition, temperature, and density.  

For steady-state accretion, we have a constant flow of mass, $\dot{m}$, through each shell, and the solution is time-independent.  We can then replace the Lagrangian time derivatives by total derivatives with respect to the mass coordinate, $\mathrm{d}m=4\pi{}r^2\rho\,\mathrm{d}r$, using 
\begin{align}
\pdv{}{t} \rightarrow \dot{m}\frac{\mathrm{d}}{\mathrm{d}m}
\;.
\end{align}
We neglect the acceleration term in Eq.~(\ref{eq:momentum}).  
Formally, the accretion flow advection velocity could be written as $$
\frac{\mathrm{d}r}{\mathrm{d}t}=-\frac{\dot{m}}{4\pi{}r^2\rho}
$$ 
and the resulting (stationary) acceleration is $$
\frac{\mathrm{d}^2r}{\mathrm{d}t^2}=\frac{\dot{m}^2}{4\pi{}r^2\rho^2}\frac{\mathrm{d}\rho}{\mathrm{d}m}\;.
$$  These accelerations are very small compared to the gravitational acceleration. 

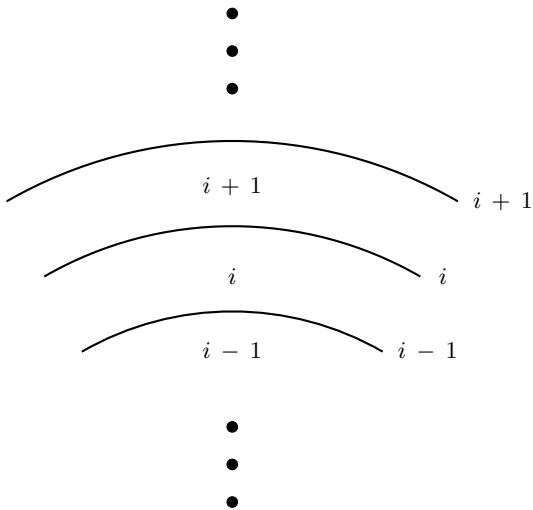
\begin{figure}
    \centering
    \begin{tikzpicture}
        \filldraw[black] (-3,2.5) circle (2pt);
        \filldraw[black] (-3,2) circle (2pt);
        \filldraw[black] (-3,1.5) circle (2pt);
        \draw[black, thick] (0,0) arc (60:120:6);
        \draw[black, thick] (-.5,-1) arc (60:120:5);
        \draw[black, thick] (-1,-2) arc (60:120:4);
        \filldraw[black] (-3,-3) circle (2pt);
        \filldraw[black] (-3,-3.5) circle (2pt);
        \filldraw[black] (-3,-4) circle (2pt);
        \node[align=center, text width=3cm] at (-3,.2) {$i+1$};
        \node[align=center, text width=3cm] at (-3,-1) {$i$};
        \node[align=center, text width=3cm] at (-3,-2) {$i-1$};
        \node[align=center, text width=3cm] at (.6,0) {$i+1$};
        \node[align=center,text width=3cm] at (-.2,-1) {$i$};
        \node[align=center, text width=3cm] at (-.4,-2) {$i-1$};
    \end{tikzpicture}
    \caption{A schematic diagram shows how the stellar structure is constructed by isotropic zones for simulations.  Time steps and the resolutions of variables are determined according to the zone sizes. The grid variable for generic zones naming scheme is defined as follow:\\
    \smallskip
    \begin{tabular}{ll}
    \hline
    \noalign{\smallskip}
    $i+2$ & two zones above \\
    $i+1$ & one zone above \\
    $i$ & current zone \\
    $i-1$ & one zone below \\
    $i-2$ & two zones below \\
    \noalign{\smallskip}
    \hline
    \end{tabular}
    }
    \label{fig:zone}
\end{figure}

The above equations are then reduced to a system of ordinary differential equations,
\begin{align}
        \dv{P}{m} &= -\frac{Gm}{4\pi r^4} \;,\label{hydrostatic_eq} \\
    \dv{l}{m} &= \epsilon_n - \epsilon_v - \dot{m}\left(\dv{u}{m} + P\dv{v}{m}\right) \;,\label{energy_eq}\\
    \dv{X_j}{m} &= 
    \dot{m}^{-1}\,F(T, \rho, X_j)\;\label{nuclear_network_eq},
\end{align}
which are independent of time, and can be solved as a boundary value problem, given boundary conditions on surface luminosity, radius, total mass, accretion composition and accretion rate.  Whereas for 1D multi-zone simulations, we numerically solve Equations~(\ref{hydrostatic_eq})--(\ref{nuclear_network_eq}) via a finite difference scheme with second order upwind discretisation.  The structure of the accreting envelope is divided into spherical shells (see the configuration in Figure~\ref{fig:zone}). Each zone is defined such that the mass increases towards the centre, but the increment may also be adjusted in order to ensure sufficient resolution for composition and thermodynamic variables.

We will go through the derivation of the discretisation of the surface, first zone, as well as generic zones in the following subsections. The descriptions of variables and constants used for \textsc{StarShot} throughout the content can be found in Table~\ref{quantity_table} if not stated otherwise.   %Each of them come with slightly different conditions.

\subsection{Outer boundary condition}
With the plane-parallel atmosphere model, we can define the surface, or photosphere in other words, located at an optical depth $\tau = \nicefrac23$ (\citealt{cox}, Section 20.1). For our models, the surface luminosity $L_\mathrm{s}$, neutron star radius $R_\star$ and mass $M$ are given. The surface temperature is obtained by the relation 
\begin{equation}
    \label{lum}
        L_{\text{s}} = 4\pi R_\star^2\sigma T_{\text{s}}^4\;.
\end{equation}
 Under the conditions of hydrostatic equilibrium, we have the relation at the surface
\begin{align}
    \dd P &= \frac{GM\rho_{\text{s}}}{R_\star^2} \dd r \nonumber\\
    \label{hydro_op}
    &= \frac{GM}{\kappa_{\text{s}} R_\star^2} \dd \tau\;,
\end{align}
if we expressed $\mathrm{d}r=\mathrm{d}\tau/\kappa\rho$. Integrating Equation~(\ref{hydro_op}) from $\tau=0$, where we assume the pressure approximately goes to zero, to the surface at $\tau = \nicefrac23$, we obtain
\begin{equation}
    \label{P_s}
        P_\text{s} = \frac{2}{3}\frac{GM}{\kappa_{\text{s}} R_\star^2}\;,
\end{equation}
where $\kappa_{\text{s}}(T_{\text{s}}, \rho_{\text{s}})$ is computed by the \textsc{Kepler} \texttt{eos} module. It takes radiative opacity \citep{radiative_opac} and electron conductivity \citep{opacity_paper}  into account. $P_\text{s}$ is sustained by the radiation and gas pressure from the surface boundary, i.e., 
\begin{equation}
    \label{hydro_equ_surface}
    P(T_{\text{s}}, \rho_{\text{s}}) = P_{\text{rad}}(T_{\text{s}}) + P_{\text{gas}}(T_{\text{s}}, \rho_{\text{s}}) = P_\text{s} \;,
\end{equation}
where $P(T_{\text{s}}, \rho_{\text{s}})$ is also computed by the \textsc{Kepler} \texttt{eos} module, which allows for a general mixture of radiation, ions, electrons and relativity \citep{1978ApJ...225.1021W}. On top of that, the module also returns internal energy of gas, along with the partial derivatives of all returned variables with respect to temperature and density respectively. 

Since the composition of the accreted gas is specified, and the surface temperature is found through Equation~(\ref{lum}), the only variable required to solve Equation~(\ref{hydro_equ_surface}) is the density for the surface.  We used Newton-Raphson reduction to numerically solve it in iterations, in which the approximation for the variable is
\begin{equation}
    \label{newton_raphson_1d}
        \rho_{\text{s}} \prime = \rho_{\text{s}} - \left(P - \frac{2}{3}\frac{GM}{\kappa_{\text{s}} R_\star^2} \right)\, \left(\dv{P}{\rho_{\text{s}}} +  \frac{2}{3}\frac{GM}{\kappa_{\text{s}}^2 R_\star^2}\dv{\kappa_{\text{s}}}{\rho_{\text{s}}}\right)^{\!-1}\;, 
\end{equation}
where ${\rho_{\text{s}}}\prime$ is the refined density.  We stop iterating once once Equation~(\ref{hydro_equ_surface}) converges within a given tolerance. 
 $\rho_{\text{s}}$, along with $T_{\text{s}}$ are used as the initial guesses for the next zone.  The composition of accretion also reserves for the next zone as no nuclear burning is considered.

For the next (outermost) zone, the centre is defined at half of its mass.  The hydrostatic equilibrium for this zone has an extra term compared to Equation~(\ref{hydro_equ_surface}). This additional pressure is contributed by the mass accumulated from the surface to the first zone centre, i.e.,
%By discretising Equation~(\ref{hydrostatic_eq}), the pressure between two adjacent shells can be related as
\begin{equation}
    \label{hydro_1stzone}
        P(T_N, \rho_N, X_N) = P_{\text{s}} + \frac{G\Delta m_N}{8\pi R_\star^2} = P_N\;,    
\end{equation}
where $N$ is the index for the outermost zone.  Radiative diffusion of energy is also considered from this zone on, adding another condition for the variables;
\begin{equation}
    \label{luminosity}
    L(T_N, \rho_N) = -\frac{4\pi acR_\star^2}{3\rho_N \kappa_N}\frac{T_N^4 - T_{\text{s}}^4}{r_{\text{c}(N)} - R_\star} = L_N = L_{\text{s}}\;,
\end{equation}
where $L_N$ reserves the luminosity at the surface.  Equation~(\ref{luminosity}) represents the radiative diffusion from the first zone centre ($r = r_{\text{c}(N)}$) to the surface boundary $(r = R_\star$), crossing half of the zone. The former radius is expressed as:
\begin{equation}
    r_{\text{c}(N)} = \sqrt[3]{\frac{R_\star^3 - 3\Delta m_N}{8\pi \rho_N}}\;.
\end{equation}
In addition to density, temperature now becomes another independent variable to solve the hydrostatic equilibrium (Equation~\ref{hydro_1stzone}), along with the extra energy equation (Equation~\ref{luminosity}), compared to the surface. Therefore, the new approximation for variables by Newton Raphson method becomes two dimensional, i.e.,
\begin{equation}
    \begin{bmatrix}
        T'_N \\ \rho'_N
    \end{bmatrix}
    =
    \begin{bmatrix}
        T_N \\ \rho_N
    \end{bmatrix}
    -
    \begin{bmatrix}
        c_1 \\ c_2
    \end{bmatrix}\;,
\end{equation}
where the last term is the solution of the linear system
\begin{equation}
    \begin{bmatrix}
        \pdv{P}{T_N} & \pdv{P}{\rho_N} \\ \pdv{L}{T_N} & \pdv{L}{\rho_N} 
    \end{bmatrix}
    \begin{bmatrix}
        c_1 \\ c_2
    \end{bmatrix}
    =
    \begin{bmatrix}
        P - P_N \\ L - L_N\;
    \end{bmatrix}\;,
    \label{linear}
\end{equation}
which represents the corrections for the two variables. They are accepted once both Equation~(\ref{hydro_1stzone}) and~(\ref{luminosity}) converge, and being used as the initial values for the next zone. Furthermore, the composition of accretion reserves for the next zone again as we do not consider nuclear burning here.

\subsection{Generic zones}
\label{generic_zone}
\begin{figure*}
    \centering
    \begin{tikzpicture}[node distance=2cm]
        \tikzstyle{r} = [rectangle, rounded corners, minimum width=3cm, minimum height=1cm,text centered, text width=4cm, draw=black]
        \tikzstyle{rec2} = [rectangle, rounded corners, minimum width=3cm, minimum height=1cm,text centered, text width=3.9cm, draw=black]
        \tikzstyle{arrow} = [thick,->,>=stealth]        
        \node (r0) [r] {Initial $T_i, \rho_i$};
        \node (r1) [r, below of=r0] {Define zone mass};
        \node (r2) [r, below of=r1] {Compute variables and 
        update abundance of elements };
        \node (r3) [r, below of=r2] {Errors of Eq.~(\ref{li_li+1}) and~(\ref{hydro_generic}) $<$ threshold};
        \node (r5) [r, below of=r3] {Check whether changes of abundance are too large};
        \node (r6) [rec2, below of=r5] {Accept variables and move to the next zone};
        \coordinate (Middle) at ($(r3)!0.5!(r2)$);
        \node (r7) [r, right of=Middle, node distance=5cm] {Newton-Raphson system reduction};
        \node (r8) [r, left of=Middle, node distance=5cm] {Zone mass reduction};        
        \draw [arrow] (r0) -- (r1);
        \draw [arrow] (r1) -- (r2);
        \draw [arrow] (r2) -- (r3);
        \draw [arrow] (r3) -- (r5);
        \draw [arrow] (r3) -- node[anchor=east] {yes} (r5);
        \draw [arrow] (r5) -- node[anchor=east] {pass} (r6);
        \draw [arrow] (r3) -| node[below ,near start] {no} (r7);
        \draw [arrow] (r7) |- node[above, near end] {$T'_i
        \;, \rho'_i$} (r2);
        \draw [arrow] (r5) -| node[below ,near start] {fail} (r8);
        \draw [arrow] (r8) |-  (r1);
        \draw [dashed, arrow] (r6) -- ++(8,0) |-  (r0);
    \end{tikzpicture}
\caption{Algorithm for generic zones.  The dashed arrow represents proceeding to the next zone, whereas the two inner loops with solid arrows are the main process of the computation of variables.  Within the most inner loop, temperature and density are being refined through Newton-Raphson reduction.  Other variables are recomputed with the updated temperature and density accordingly in each iteration.  Once the errors of Equation~(\ref{li_li+1}) and~(\ref{hydro_generic}) is less than the threshold, the loop is exited and the code proceeds to the next step to determine whether the changes of abundances are too large. If it fails, the code redoes the above steps with smaller zone mass via the outer solid loop.  Otherwise, the calculation of the current zone is then finalised.  The code then proceeds to the next zone (via dashed arrow) where the resolved temperature and density from the previous zone are used to be the initial values.}
\label{flowchart}
\end{figure*}
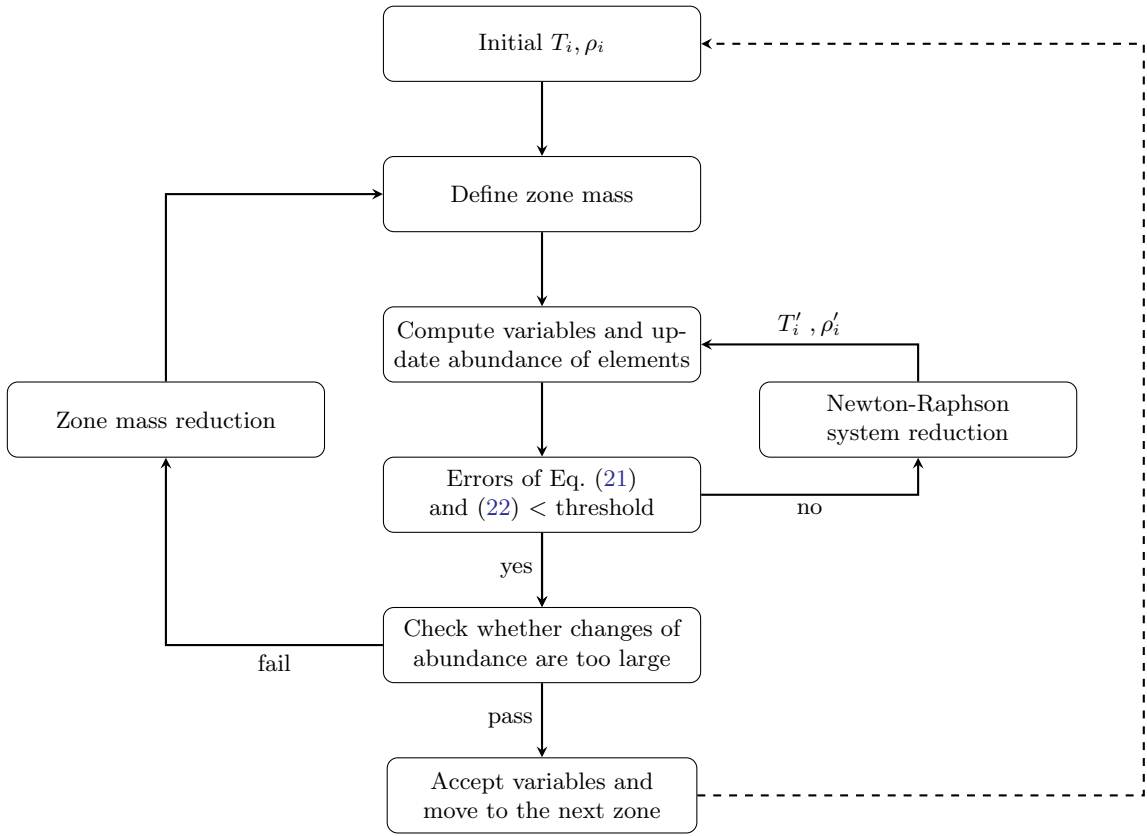

The calculations for the next step applied to the remaining zones, where nuclear reactions are also taken into account (see Sec.~\ref{nuclear} for the employed nuclear networks).  We start applying indices, explained in Figure~(\ref{fig:zone}), to identify the zones being referenced.  Advection of accreted gas is also included for energy conservation. To deal with this energy term caused by the movement of gas, we consider the time for gas to flow from the centre of Zone $i+2$ to $i+1$, which is equal to $(\Delta m_{i+2} + \Delta m_{i+1})/(2\dot{m})$. For pressure at this intermediate boundary, we defined it as a harmonic mean of the pressure at the two zones:
\begin{equation}
    P_{\text{b}(i+1)} = \frac{2P_{i+2}P_{i+1}}{P_{i+2}+P_{i+1}}\;.
\end{equation}
Hence, the change of specific enthalpy per unit time caused by advection at Interface $i+1$ can be written as:
\begin{align}
    \label{pdv}
        w_{i+1} &= \Delta u + P_{\text{b}(i+1)}\Delta v \nonumber\\
           &= \left[u_{i+2} - u_{i+1} + \frac{2P_{i+2}P_{i+1}}{P_{i+2}+P_{i+1}}\left(\frac{1}{\rho_{i+2}} - \frac{1}{\rho_{i+1}}\right)\right] \notag \\ 
           &\times \frac{2\dot{m}}{(\Delta m_{i+2} + \Delta m_{i+1})}\;.
\end{align}
The same derivation can be applied to Interface $i$ to obtain $w(T_i, \rho_i)$. 
 Therefore, the net energy rate including nuclear reactions for Zone $i+1$ equals to 
\begin{equation}
    \label{total energy rate} 
        E_{i+1} = \left(\frac{w_{i+1} + w(T_i, \rho_i)}{2} + \epsilon_{n(i+1)} - \epsilon_{v(i+1)} \right)\Delta m_{i+1},
\end{equation}
where the factor $\nicefrac{1}{2}$ comes from averaging the boundary values for the zone centre. The energy conservation gives the relation of the luminosity at interfaces $i$ and $i+1$ to be
\begin{equation}
    \label{li_li+1}
          L(T_i, \rho_i)  = -\frac{4\pi acr_i^2}{3\rho_{\text{b}i}\, \kappa_{\text{b}i}}\frac{T_{i}^4 - T_{i+1}^4}{r_{\text{c}(i)} - r_{\text{c}(i+1)}}
              = L_{i+1} - E_{i+1} = L_i, 
\end{equation}
where $L_{i+1}$ is independent of $T_i$ and $\rho_i$.  The hydrostatic equilibrium condition Equation~(\ref{hydro_1stzone}) is slightly modified to  
\begin{align}
    \label{hydro_generic}
        P(T_i, \rho_i, X_i) = P_{i+1} + \frac{(\Delta m_i+\Delta m_{i+1}) g_i}{8\pi r_i^2} = P_i\;,     
\end{align}
considering the accumulative pressure from half of the mass from both Zone $i+1$ and $i$ (centre to centre).  The linear system of the saver becomes 
\begin{align}
    \begin{bmatrix}
        \pdv{P}{T_i} & \pdv{P}{\rho_i} \\ \pdv{L}{T_i} & \pdv{L}{\rho_i} 
    \end{bmatrix}
    \begin{bmatrix}
        c_1 \\ c_2
    \end{bmatrix}
    =
    \begin{bmatrix}
        P - P_i\\ L - L_i\;
    \end{bmatrix}\;,
    \label{linear_generic}
\end{align}
where the partial derivatives in the first row of the Jacobian matrix are computed by the Kepler \texttt{eos} module, while the details of the ones in the second row can be found in Appendix~\ref{append}.  Meanwhile, abundances of elements are updated in each iteration by the nuclear networks, which also results in changes of opacity for consistency. Once both Equations~(\ref{li_li+1}) and~(\ref{hydro_generic}) converge with the approximated variables, there is a final check done by the networks for mass conservation that prevents abrupt changes of the abundances of elements, and therefore keeping reasonable resolution for variables.  If failed, the calculation of the current zone is repeated with a smaller zone mass.  Otherwise, the next zone is proceeded.  We summarise the algorithm for generic zones in Figure~(\ref{flowchart}). 
% Finally, at the end of each run, the code will truncate a zone mass in order to have the penultimate boundary located at the designated depth, where base luminosity is defined. 

% Quantity table
\begin{table}
\centering
\caption{Variables and Constants used for \textsc{StarShot}}
\label{quantity_table}
% \begin{threeparttable}
\begin{tabular}{lcl} % four columns, alignment for each
\hline\hline
Location & Name & Description\\
\hline
On zone & $P$ & Pressure\\
& $T$ & Temperature\\
& $\rho$ & Density\\
& $\kappa$ & Opacity\\
& $u$ & Specific internal energy of gas\\
& $v$ & Specific volume\\
& $\Delta m$ & zone mass\\
& $r_{\text{c}}$ & Radius\\
& $E$ & Energy rate\\
& $\epsilon_\text{n}$ & Specific nuclear energy\\ 
& & generated rate\\
& $\epsilon_\nu$ & Specific neutrino loss rate\\
\hline
At interface
& $L$ & Luminosity\\
& $r$ & Radius\\
& $g$ & Local gravity\\
& $w$ & Specific enthalpy\\
\hline
Constant
& $\sigma$ & Stefan-Boltzmann constant\\
& $c$ & Speed of light\\
& $M$ & Mass of neutron star\\
& $R\star$ & Radius of neutron star\\
& $G$ & Gravitational constant\\
& $\dot{m}$ & Accretion rate\\
\hline\hline
\end{tabular}
\begin{tablenotes}
\item{\textsc{NOTE}: In addition, we use the subscript `s' to represent surface values, `$N$' to the first zone, and `b' to define boundary values for on zone variables, e.g., $P_{\text{b}}\,, \rho_{\text{b}}$, etc., throughout the content.}
\end{tablenotes}
% \end{threeparttable}
\end{table}

\subsection{The Employed Nuclear Network}
\label{nuclear}
We use the first-order implicit adaptive nuclear reaction network from \textsc{Kepler} \citep{2002ApJ...576..323R}, which includes all nuclear species and reactions up to astatine except fission.  
Due to the potentially large number of isotopes that may need to be added from zone to zone, we developed a new \emph{iterative} adaptive network that repeats a timestep until there is no more addition of new species.  This is an extension beyond the current \textsc{Kepler} network that only adjusts the network at the end of a time step in preparation for the next time step.  For most stellar applications that are time-dependent and resolve change of abundances with time for Lagrangian zones, this approach is usually sufficient.  The application of \textsc{StarShot} to X-ray burst (XRB) simulations, however, can encounter sharp transitions between adjacent shells.  One test example, e.g., was the slow accretion of pure $^1$H onto cold neutron star.  In this case, when the threshold for electron capture on protons is reached, heavy nuclei with $A>100$ may suddenly form over a small range of column depths, basically almost from one zone to the next, with the nuclear reaction flow bypassing lighter nuclei.  Additionally, we iteratively sub-cycle the nuclear reaction network calculation when the abundance changes are too large or when mass conservation is violated by more than one in $10^{-14}$.  The same framework is also used for the \textsc{Burn} code in \citet{Jakobus22}. 

% \textsc{StarShot} employs \textsc{Kepler} net module uses the JINA-CEE reaclib Version 2.2 with more than 6,000 isotopes \citep{cyburt2010} to estimate nuclear reaction rates . Nuclear reactions are not taken into account for the surface and first zones. 

% \subsection{Adaptive Zonal Step Size}
% \label{step_size}
% The first zone under the surface is defined in mass $10^{13}\,\mathrm{kg}$, while the subsequent zones are defined by $1.2$ times more massive than the previous one. As it goes into deeper layers, the mass difference between two consecutive zones becomes larger which may result in low resolution and discontinuities of stellar structures. In light of that, we trace the changes of abundance for all isotopes when creating a new zone. If any of them exceeds a fractional change of $10^{-3}$, the new zones is redefined with a smaller mass. This process is repeated until the condition being met. 

\section{Results}
\label{sec:results}
\begin{figure*}[t!]
\centering
\subfigure[H-model]{
\includegraphics[width=\columnwidth]{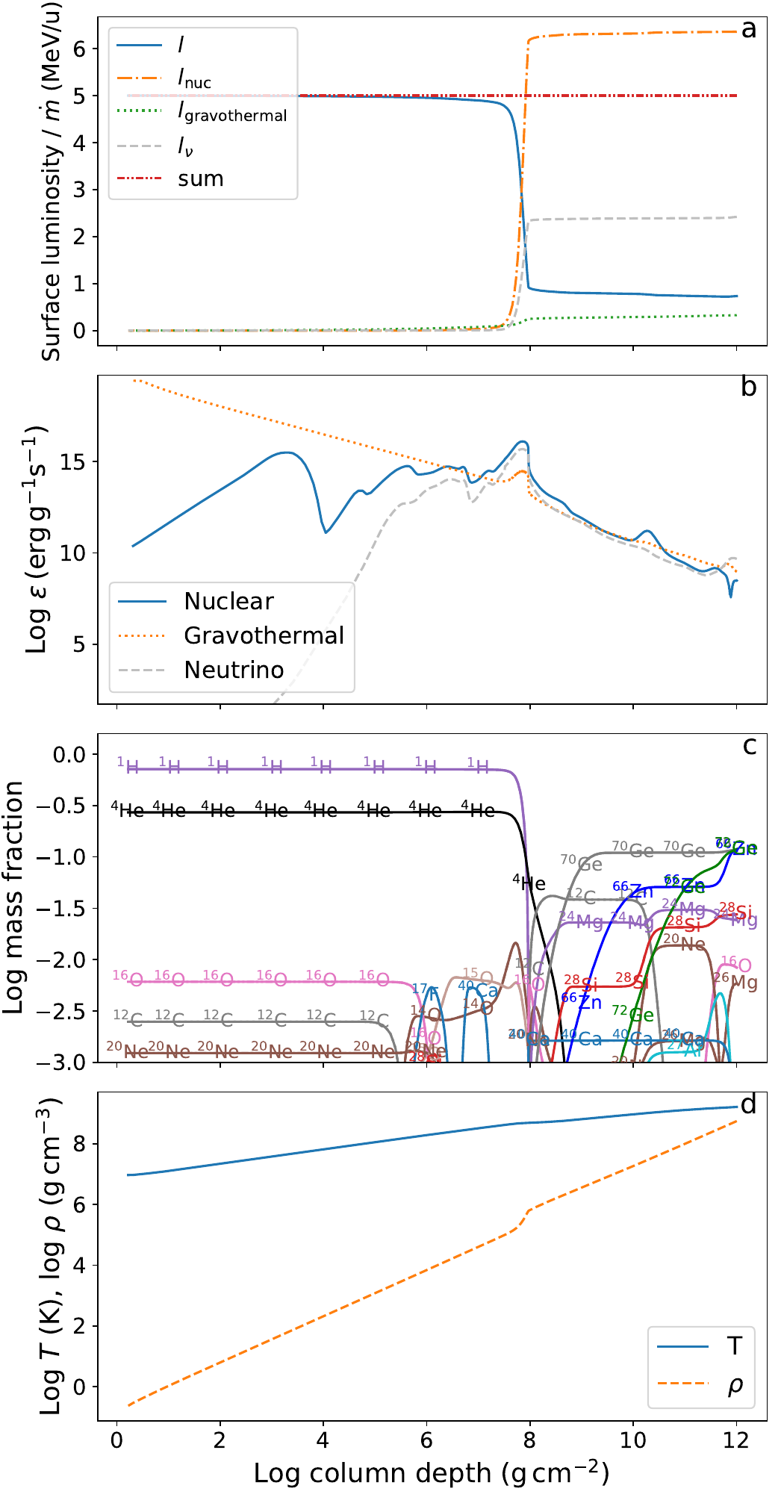}
}
\centering
\subfigure[He-model]{
\includegraphics[width=\columnwidth]{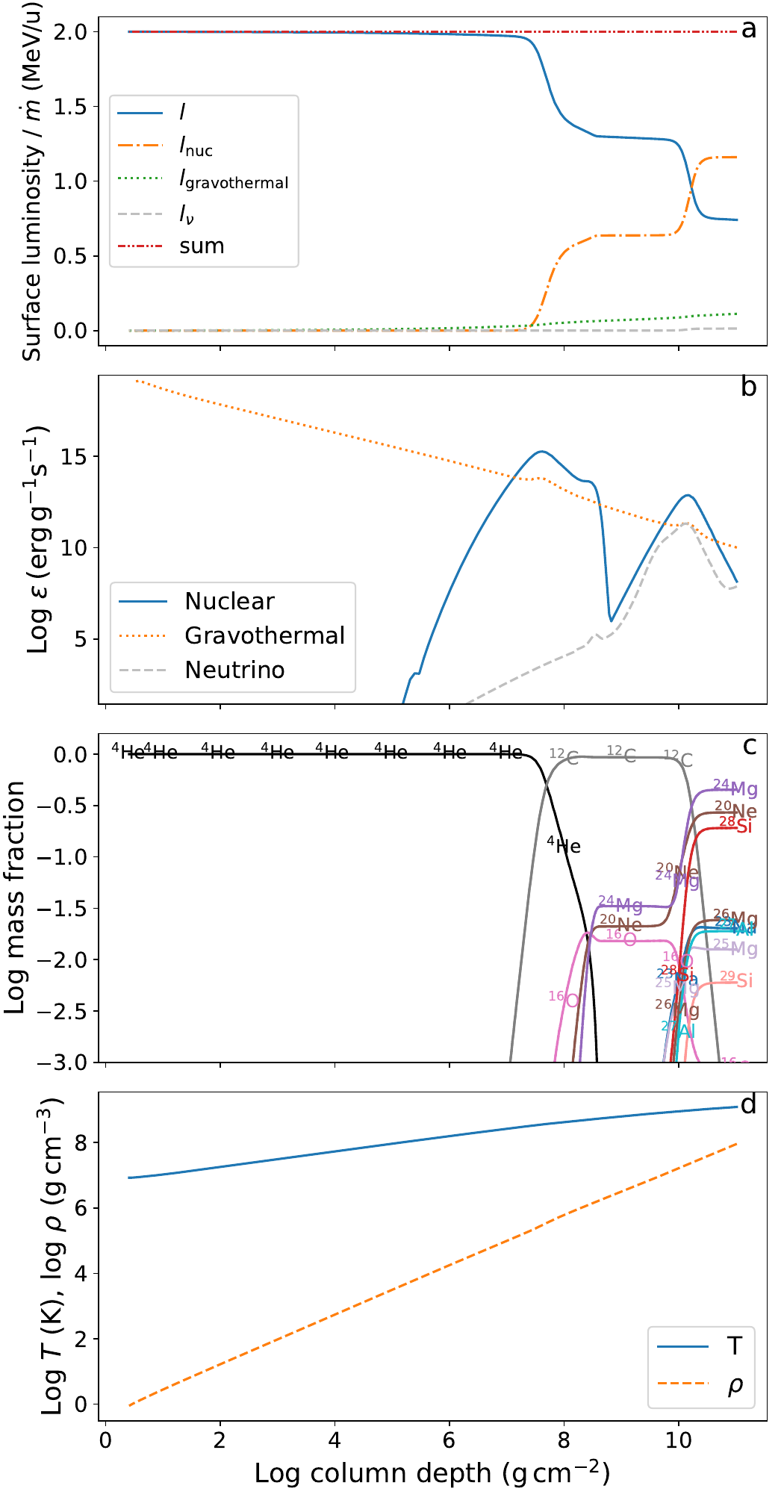}
}
\caption{Combined plots for solar abundance accretion model (\textsl{H-model}) and pure helium model (\textsl{He-model}).  Input parameters can be found in Table~\ref{quantity_table}.  We assume the neutron stars having $M = 1.4\,M_{\odot}$ and $R = 11\,\mathrm{km}$.  Both runs accrete materials to a depth at $10^{11}\,\mathrm{g\,cm}^{-2}$.  Panel a shows the integrated flux ($l$) contributed by nuclear energy ($l_{\text{nuc}}$), neutrino loss ($l_{\nu}$) and gravothermal energy ($l_{\text{gravothermal}}$), in the unit of $\textrm{MeV}$ per accreted nucleons; Panel b, the specific energy generated rate from nuclear reaction, neutrino loss and gravothermal; Panel c, the evolution of abundances (elements up to mass number $=50$ and the top three most abundant at the base for the H model); and Panel d, the temperature and density. All the panels are plotted as a function of column depth.}
\label{plot}
\end{figure*}

In this section, we show two examples of \textsc{StarShot} with their corresponding parameters listed in Table~\ref{runs}. Both runs accrete to a column depth at $10^{11}\,\mathrm{g\,cm}^{-2}$. We use $M=1.4\,M_{\odot}$ and $R_\star=10\,\mathrm{km}$ for the neutron star mass and radius respectively, which gives the GR correction of surface gravity of the same mass but an $11.2\,\mathrm{km}$ radius counterpart \citep{keek2011}. Each result comes with a plot of four panels (Figure~\ref{plot}). 

\begin{figure}
	\includegraphics[width=\columnwidth]{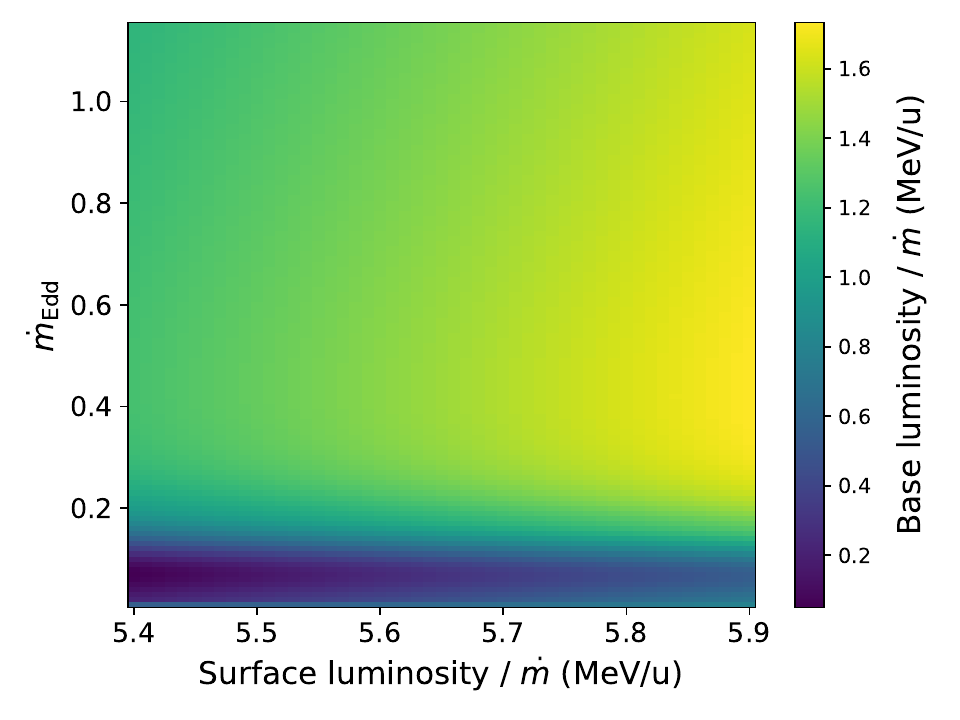}
    \caption{Base flux at column depths $10^{11}\,\mathrm{g}\,\mathrm{cm}^2$ as a function of surface luminosity and accretion rate.  Both, base flux and surface luminosity are provided as energy per accreted nucleon; the actual luminosity and flux then scale with the accretion rate.  The plot has total $51$ (x-axis) $\times\;115$ (y-axis) runs. 
 For an energetically neutral layer, one would expect that the base flux scales linearly with the accretion luminosity.  Depending on the thermodynamic conditions, the different burning regimes cause different amounts of nuclear energy release, resulting in variations of the base flux.  On top of that, the amounts of gravitation potential energy release are different due to the more or less extended envelopes at the same column depth, which also contributes to the variations.  That is, for different accretion rates, the same base flux can result in different surface luminosities per accreted nucleon.  Note that our model is for the stable, stationary solution. In nature, however, there are many accretion regimes in which the stratification may be unstable to hydrodynamic and thermonuclear instabilities, which would have significant impact on the corresponding base fluxes.  Note that our model is for the stable, stationary solution. In nature, however, there are many accretion regimes in which the stratification may be unstable to hydrodynamic and thermonuclear instabilities, which would have significant impact on the corresponding base fluxes.
    }
    \label{fig:qqb}
\end{figure}

The integrated luminosity, provided as energy per accreted nucleon, versus column depth is shown in Panel a. The blue curve represents the total luminosity at local interfaces.  As a boundary condition, it integrates towards the centre of the star with accumulated nuclear energy heating rate ($l_{\text{nuc}}$), loss via neutrino ($l_{\nu}$) and energy contributed by advection ($l_\text{gravotherm}$).  To estimate the base luminosity, \textsc{StarShot} solves boundary value problems for the entire grid of zones, one by one from the surface all the way to the designated column depth, which equals to $10^{11}\,\mathrm{g\,cm}^{-2}$ for our results.  We also show a grid of simulations to estimate the base luminosity as a function of surface luminosity and accretion rate with solar composition (See Figure~\ref{fig:qqb}).

The local specific nuclear energy generation rate, neutrino loss rate and gravothermal energy release rate versus the column depth is shown in panel b, whereas panel c shows the evolution of abundances over depth.  For \textsl{H-model}, only the top three most abundant at the base, as well as elements up to mass number $=50$ are displayed. Compared to He ignition, H burns through CNO cycle in shallower layers, as  shown by the production of $^{14}$O and $^{15}$O in the \textsl{H-model}. At this stage, helium still remains abundant before the triple$-\alpha$ reaction starts setting in. This reaction takes place at a deeper depth, which is indicated by the production of $^{12}$C in the \textsl{He-model}.  At the depth $\sim 10^8\,\mathrm{g\,cm}^{-2}$, the breakout reaction of $\beta$-CNO cycles leads to alpha capture processes, creating elements in the iron group.  They further burn to heavier elements via the \textsl{rp} process. Therefore, both H and He levels drop significantly at this depth associated with the highest nuclear energy rate being generated, as shown in the \textsl{H-model}.  In the pure-helium accretion model, massive carbon, which is believed to be the fuel of superbursts \citep{cumming_superburst}, is created in the NS shallow ocean (from $10^8$ to $10^{10}\,\mathrm{g\,cm}^{-2}$ column depth). It can be immediately destroyed through both $^{12}\mathrm{C}(p,\gamma)^{13}\mathrm{N}$ and $^{12}\mathrm{C}(\alpha,\gamma)^{16}\mathrm{O}$ reactions if any of the fuel presents \citep{carbon_synthesis}. The limited hydrogen, as well as the vast production of carbon by triple$-\alpha$ process keep the level of its abundance high compared to the \textsl{H-model}, although it is also being consumed by the $\alpha$ capture process that creates $^{16}$O.  At the depth $\sim 10^{10}\,\mathrm{g\,cm}^{-2}$, the temperature and density are high enough to ignite carbon burning. Therefore, one of the reaction products, $^{24}$Mg, becomes abundant.  On top of that, the level of $^{12}$C drops, associated with a bump of energy in panel b for both models.

The temperature and density profile versus column depth can be found in the bottom panel. Both models show flatter temperature gradient in depth $\gtrsim 10^8\,\mathrm{g\,cm}^{-2}$.  The sharp rise of density in the \textsl{H-model} comes from the increase of mean molecular weight at the depth where most of the hydrogen burns to heavier elements. 

\begin{table*}[t!]
	\centering
	\caption{Parameters for \textsc{starshot} runs}
	\begin{threeparttable}
	\begin{tabular}{lcccl} % four columns, alignment for each
		\hline\hline
		Model &Input surface luminosity ($\mathrm{MeV/u}$)&  $\dot{M}_{\mathrm{EDD}}$ &Acc. Composition &  Output base luminosity ($\mathrm{MeV/u}$)\\
		\hline
		H & 5.0 & 1 & solar\textsuperscript{a} & 0.73
		\\
		He & 2.0 & 1 & pure helium & 0.68
		\\
		\hline
	\end{tabular}
	\begin{tablenotes}
	\item[a] We use solar abundances from \cite{SolAbu}
	\end{tablenotes}
	\end{threeparttable}
	\label{runs}
\end{table*}

% The base flux scales linearly with the accretion luminosity, and shows a similar relation with the accretion rate, except for the low accretion regimes.  For higher accretion rates, with more fuels to burn, envelope gets hotter and less denser such that it is more extended to reach the same column depth.  The amounts of gravitation potential energy release are therefore higher at the same column depth, which in turn is the dominant factor in the variations of the base flux.  There are also other factors contributing to the variations, such as nuclear energy release or neutrino loss that depend on the varying thermodynamic conditions.

% For an energetically neutral layer, one would expect that the base flux scales linearly with the accretion luminosity, but the different burning regimes cause different amounts of nuclear energy release depending on the varying thermodynamic conditions, resulting in variations of the base flux.  That is, for different accretion rates, the same base flux can result in different surface luminosities per accreted nucleon.  Note that our model is for the stable, stationary solution. In nature, however, there are many accretion regimes in which the stratification may be unstable to hydrodynamic and thermonuclear instabilities, which would have significant impact on the corresponding base fluxes.

\section{Summary and outlook}
We introduce our \textsc{StarShot} code which, combined with the \textsc{Kepler} \texttt{eos} module and nuclear networks, implements time-independent simulations for steady-state accretion onto compact objects.  Compared to traditional stellar evolution codes, the main difference of \textsc{StarShot} is that the full set of differential equations (Equations~\ref{eq:momentum}--\ref{tradition_eqs}) are reduced to be decoupled with time (Equations~\ref{hydrostatic_eq}--\ref{nuclear_network_eq}).  With that, the computational cost for stable burning simulations is reduced from days to minutes.  Therefore, it is able to resolve nuclear burning with unprecedented resolution, and is able to explore deeper layers of neutron stars.

We go through the details of the mathematical derivation in discretised manner for the surface, first and generic zones individually as they involve different stellar conditions.  Furthermore, we summarise the algorithm for the calculations of generic zones.  The code also includes adaptive step size adjustment to prevent from substantial changes in chemical abundances over the entire grid of simulations.  We show two examples of \textsc{StarShot} with steady-state H-rich and pure helium accretion onto a neutron star respectively, as well as the estimation of base luminosity with combinations of surface luminosity and accretion rates.

One direct application of the code is to explore stable burning regimes, where substantial carbon is produced closed to the ignition depth of superbursts at $\sim 10^{12}\,\mathrm{g\,cm}^{-2}$.  \citet{keek2016} predicted that a stable burning regime of solar composition material at close to 10 percent Eddington accretion rate could produce as much as 98\% carbon for the ignition of superbursts.  To continue the study, one can use \textsc{StarShot} to easily generate grids of models to explore the chemical composition of the neutron star outer crust.

Another application is to provide initial conditions for dynamical codes. The $\mathrm{mHz}$ quasi-periodic oscillations (QPOs) found in LMXBs are thought to be marginally stable nuclear burning at high accretion rates \citep{paczynski1983}.  Eight sources have been found showing these signals during outburst \citep{tse2021}.  \citet{heger_qpo} was able to reproduce this special periodic oscillations using the stellar evolution code \textsc{Kepler}.  At the beginning of each the simulations, an initial condition that closed to stability boundary has to be achieved.  We intend to use \textsc{StarShot}, rather than waiting for the dynamical code, to generate initial conditions, which are then employed to time-evolution codes and allow perturbation to continue the simulations.

Last but not least, decades of research have been done on the thermal stability analysis of nuclear burning on LMXBs (see, for example,  \citealt{paczynski1983,Narayan2003,zamfir2014}).  Previous studies of using steady-state codes for thermal stability analysis are limited to either single zone models or incomprehensive nuclear networks.  For time-evolution codes, there are also attempts to explore stability boundaries, but they were only able to cover small regions over the whole parameter space, due to the expensive computational time.  To better handle the aforementioned challenges, a possible future extension is to incorporate thermal stability analysis to \textsc{StarShot} to continue the study, which takes the advantages of multizone calculations, sophisticated nuclear networks and efficient computational time.

\bibliography{ref}{}
\bibliographystyle{aasjournal}

\appendix
\section{Derivation of partial derivatives}
\label{append}
In Section~\ref{generic_zone}, we show a two-dimensional linear system for the approximation of variables in Equation~(\ref{linear_generic}).  The partial derivatives in the first row of the Jacobian are computed by the \textsc{Kepler} \texttt{eos} module.  To derive the ones in the second row, we will first show an alternative expression of luminosity $L$, followed by its partial derivatives with respect to the temperature and density, $T_i$ and $\rho_i$, individually.  Then, we derive the partial derivatives of energy rate, $E_{i+1}$ from Equation~(\ref{total energy rate}).  The descriptions of variables can be found in Table~\ref{quantity_table}, whereas the meaning of indexes can be referred to Figure~(\ref{fig:zone}). 

From Equation~(\ref{li_li+1}), 
\begin{equation}
    \label{deviation_of_lumi}
    L(T_i, \rho_i) = -\frac{4\pi acr_i^2}{3\rho_{\text{b}i}\, \kappa_{\text{b}i}}\frac{T_i^4 - T_{i+1}^4}{r_{\text{c}(i)} - r_{\text{c}(i+1)}}\;. 
\end{equation}
Using the fact that
\begin{align}
    \rho_{\text{b}i} = \left. \frac{\Delta m_i + \Delta m_{i+1}}{2}\right/ \left[\frac{4}{3}\pi (r_{\text{c}(i+1)}^3 - r_{\text{c}(i)}^3) \right] \approx \left. \frac{\Delta m_i + \Delta m_{i+1}}{2}\right/ \left[4\pi r_i^2(r_{\text{c}(i+1)} - r_{\text{c}(i)}) \right]\;, \\
    \rho_{\text{b}i}(r_{\text{c}(i+1)} - r_{\text{c}(i)})  = \frac{\Delta m_i + \Delta m_{i+1}}{8\pi r_i^2}\;, \text{as well as } \frac{1}{\kappa_{\text{b}i}} = \frac{1}{2}\left( \frac{1}{\kappa _i} + \frac{1}{\kappa_{i+1}}\right)\;, 
\end{align}
we have
\begin{align}
                L(T_i, \rho_i) &= \frac{4\pi acr_i^2(T_i^4 - T_{i+1}^4)}{3}\times \frac{8\pi r_i^2}{\Delta m_i + \Delta m_{i+1}}\times \frac{1}{2}\left( \frac{1}{\kappa _i} + \frac{1}{\kappa_{i+1}}\right) \\
                    &= \frac{(4\pi r_i^2)^2ac(T_i^4 - T_{i+1}^4)}{3(\Delta m_i + \Delta m_{i+1})}\left( \frac{1}{\kappa _i} + \frac{1}{\kappa_{i+1}}\right)\;.
\end{align}
Regarding to the partial derivatives of $L_i$ and $E_i$, with respect to $T_i$ and $\rho_i$, 
\begin{align}
    \label{d-luminosity}
    \pdv{L(T_i, \rho_i)}{T_i} &= \pdv{}{T_i}\left[ \frac{(4\pi r_i^2)^2ac(T_i^4 - T_{i+1}^4)}{3(\Delta m_i + \Delta m_{i+1})}\left( \frac{1}{\kappa _i} + \frac{1}{\kappa_{i+1}}\right) \right] \\
    &= \frac{(4\pi r_i^2)^2ac}{3 (\Delta m_i + \Delta m_{i+1})}\pdv{}{T_i}\left[ (T_i^4 - T_{i+1}^4)\left( \frac{1}{\kappa _i} + \frac{1}{\kappa_{i+1}}\right) \right]
     \\
    &= \frac{4(4\pi r_i^2)^2acT_i^3}{3(\Delta m_i + \Delta m_{i+1})}\left( \frac{1}{\kappa _i} + \frac{1}{\kappa_{i+1}}\right) - \frac{(4\pi r_i^2)^2ac(T_i^4 - T_{i+1}^4)}{3\kappa_i^2 (\Delta m_i + \Delta m_{i+1})}\pdv{\kappa_i}{T_i}\;;\\
    \nonumber \\
    \pdv{L(T_i, \rho_i)}{\rho_i} &= -\frac{(4\pi r_i^2)^2ac(T_i^4 - T_{i+1}^4)}{3\kappa_i^2(\Delta m_i + \Delta m_{i+1})}\pdv{\kappa_i}{\rho_i}\;.
\end{align}
From Equation~(\ref{total energy rate}), 
\begin{equation}
     E_{i+1} = \left(\frac{w_{i+1} + w(T_i, \rho_i)}{2} + \epsilon_{i+1}\right)\Delta m_{i+1}\;.
\end{equation}
With only $w(T_i, \rho_i)$ being as the function of $T_i$ and $\rho_i$ (in reference to Equation~\ref{pdv}), we have
\begin{align}
        \pdv{E_{i+1}}{T_i} &= \frac{\Delta m_{i+1}}{2}\pdv{w(T_i, \rho_i)}{T_i} \\
        &= \frac{\Delta m_{i+1}}{2}\pdv{}{T_i} \left\{ \left[u_{i+1} - u(T_i, \rho_i) + \frac{2P(T_i, \rho_i)P_{i+1}}{P(T_i, \rho_i)+P_{i+1}}\left(\frac{1}{\rho_{i+1}} - \frac{1}{\rho_{i}}\right)\right] \times \frac{2\dot{m}}{(\Delta m_{i} + \Delta m_{i+1})}\right\}  \\
        &= \frac{\dot{m}\,\Delta m_{i+1}}{(\Delta m_{i} + \Delta m_{i+1})} \left[-\pdv{u(T_i, \rho_i)}{T_i} + \frac{2P_{i+1}^2}{(P(T_i, \rho_i) + P_{i+1})^2} \left(\frac{1}{\rho_{i+1}} - \frac{1}{\rho_{i}}\right) \pdv{P(T_i, \rho_i)}{T_i} \right]\;;\\
        \nonumber \\ 
% density, double checked        
        \pdv{E_{i+1}}{\rho_i} &= \frac{\Delta m_{i+1}}{2}\pdv{w(T_i, \rho_i)}{\rho_i} \\
        &= \frac{\dot{m}\,\Delta m_{i+1}}{(\Delta m_{i} + \Delta m_{i+1})}\times \pdv{}{\rho_i}\left[u_{i+1} - u(T_i, \rho_i) + \frac{2P(T_i, \rho_i)P_{i+1}}{P(T_i, \rho_i)+P_{i+1}}\left(\frac{1}{\rho_{i+1}} - \frac{1}{\rho_{i}}\right)\right] \\
        &= \frac{\dot{m}\,\Delta m_{i+1}}{(\Delta m_{i} + \Delta m_{i+1})}\left\{-\pdv{u(T_i, \rho_i)}{\rho_i} + \pdv{}{\rho_i}\left[\frac{2P(T_i, \rho_i)P_{i+1}}{P(T_i, \rho_i)+P_{i+1}}\left(\frac{1}{\rho_{i+1}} - \frac{1}{\rho_{i}}\right) \right]\right\} \\
        &= \frac{\dot{m}\,\Delta m_{i+1}}{(\Delta m_{i} + \Delta m_{i+1})}\left[-\pdv{u(T_i, \rho_i)}{\rho_i} + \frac{2P_{i+1}^2}{(P(T_i, \rho_i)+P_{i+1})^2}\left(\frac{1}{\rho_{i+1}} - \frac{1}{\rho_{i}}\right)\pdv{P(T_i, \rho_i)}{\rho_i} + \frac{2P(T_i, \rho_i)P_{i+1}}{P(T_i, \rho_i)+P_{i+1}}\frac{1}{\rho_{i}^2}  \right]\;.
\end{align}
\end{document}